\documentclass[a4,12pt,epsfig]{article}

\usepackage{axodraw}
\usepackage{cite}
\usepackage{latexsym} 

\begin{document}

\newcommand{\pp}{p\bar{p}}
\newcommand{\ve}{\varepsilon}           
\newcommand{\GeV}{\mbox{\rm GeV}}
\newcommand{\VS}{\rule[-2mm]{0mm}}


\newsavebox{\threediag}
\savebox{\threediag}(0,0)[bl]{
\put(-10,0){\line(1,0){10}}
\put(0,0){\line(4,3){40}}
\put(0,0){\line(4,-3){40}}
\put(0,0){\line(1,0){40}}
\put(40,30){\line(0,-1){60}}
\put(40,30){\line(1,0){10}}
\put(40,-30){\line(1,0){10}}
\put(-10,0){\vector(1,0){8}}
\put(0,0){\vector(4,3){20}}
\put(0,0){\vector(4,-3){20}}
\put(0,0){\vector(1,0){20}}
 \put(40,0){\vector(0,-1){20}}
 \put(40,0){\vector(0,1){20}}
\put(40,30){\vector(1,0){8}}
\put(40,-30){\vector(1,0){8}}
\put(0,0){\circle*{5}}
\put(40,30){\circle*{5}}
\put(40,-30){\circle*{5}}
\put(40,0){\circle*{5}}
}

\newsavebox{\twodiag}
\savebox{\twodiag}(0,0)[bl]{
\put(-10,0){\line(1,0){10}}
\put(0,0){\line(3,2){30}}
\put(0,0){\line(3,-2){30}}
\put(30,20){\line(0,-1){40}}
\put(30,20){\line(1,0){10}}
\put(30,-20){\line(1,0){10}}
\put(-10,0){\vector(1,0){8}}
\put(0,0){\vector(3,2){15}}
\put(0,0){\vector(3,-2){15}}
\put(30,20){\vector(1,0){6}}
\put(30,-20){\vector(1,0){6}}
\put(0,0){\circle*{5}}
\put(30,20){\circle*{5}}
\put(30,-20){\circle*{5}}
}

\newcommand{\FigDiag}[8]{
\mbox{
\begin{picture}(140,80)(-40,-40) 
\put(0,-30){\usebox{\threediag}} 
\put(0,0){ 
  \put(10,-20){\makebox[0mm]{#1}}
  \put(10,15){\makebox[0mm]{#2}}
  \put(25,5){\makebox[0mm]{#3}}
  \put(-20,-5){\makebox[0mm]{#6}}
  \put(46,-20){\makebox[0mm]{#4}}
  \put(46,15){\makebox[0mm]{#5}}
  \put(60,30){\makebox[0mm]{#8}}
  \put(60,-35){\makebox[0mm]{#7}}  }
\end{picture} } 
}

\newcommand{\FigDiagA}[6]{
\mbox{
\begin{picture}(110,80)(-40,-40)
\put(0,-20){\usebox{\twodiag}} 
\put(0,0){
  \put(10,-20){\makebox[0mm]{#1}}
  \put(10,15){\makebox[0mm]{#2}}
  \put(-20,-5){\makebox[0mm]{#4}}
  \put(40,0){\makebox[0mm]{#3}}
  \put(50,20){\makebox[0mm]{#6}}
  \put(50,-25){\makebox[0mm]{#5}} }
\end{picture} } 
}


\begin{flushright}
Preprint-PSI-99-31 \\
December 14, 1999 
\end{flushright}

\begin{center}
{\Large\sf
    Higher Order Two Step Mechanisms in
    Nucleon Antinucleon Annihilation
    and the OZI Rule
}\\[5mm]
{\sc  S. von Rotz$^{a,b}$, M.P. Locher$^{a,b}$ and V.E. Markushin$^a$}
\\[3mm]
{\it a) Paul Scherrer Institute, CH-5232 Villigen, Switzerland} \\
{\it b) University of Zurich, CH-8047 Zurich, Switzerland} \\
\vspace{1cm}
\end{center}

\begin{abstract}

   We evaluate three meson doorway mechanisms for nucleon-antinucleon
annihilation at rest for the first time.
   Detailed results are presented for the final state $\phi\pi^0$ 
originating from the $^3S_1$ initial state and for the
$\phi\rho$ channel originating from $^1S_0$.
   The results presented also include the improved contributions
from two meson doorway states and from the tree diagrams.  
   For all the channels considered a consistent explanation of large and 
small OZI violations emerges. 

\end{abstract}

\section{Introduction}

   Recent and accurate data for nucleon-antinucleon annihilation at rest
from experiments performed at LEAR
\cite{LEAR1,AST91,CBC93,LEAR4,CBC95,LEAR6,Ams98,OBEL98a} have challenged our
understanding of the underlying annihilation mechanisms and of the
production of mesons with hidden strangeness in particular.  Large violations
of the OZI rule for special channels have been observed.  The biggest
deviation from the OZI prediction for hadronic channels occurs for the
$\phi\pi^0$ final state and has led to speculations about the internal
structure of the nucleon suggesting a large $s\bar{s}$ component in the
wave function \cite{Ell1,Ell2,Ell3}.
   Earlier analysis \cite{2MDM0,2MDM1,2MDM2,2MDM3,2MDM4,2MDM5,2MDM6,LMR,ML98,Ma99}
has shown that two meson doorway contributions have the correct
magnitude to explain the experimental branching ratio for this reaction.
The present paper extends the preceding calculations by including three
meson doorway states.  Sizable OZI-rule avoiding contributions 
are expected from such intermediate states since the first step, 
the annihilation into three non-strange mesons, 
represents about one third of the total annihilation cross section. 

   Based on the results of \cite{SvR} we shall present a comprehensive 
effort of calculating all relevant diagrams involving (non-strange)
three meson intermediate states leading to the two meson final states 
$\phi\pi^0$ and $\phi\rho^0$.  
  The corresponding two-loop amplitudes have been evaluated with full
spin.  We have also completed the evaluation of one-loop amplitudes
(two-meson-doorways) where needed.
  For completeness we report some of the results on other two-meson 
final states.  We shall show that the results consistently explain the size 
of large and small OZI violations.

   The paper is organized as follows.  In Sec.~\ref{TMDWM} we describe
the three meson doorway formalism.  A generic case is reported in some
detail while technicalities are relegated to the Appendices.
Section~\ref{Res} presents the three-meson-doorway results.  Updated
calculations for the two-meson doorway amplitudes are included and the
full calculation is compared to the experimental branching ratios.
Section~\ref{Concl} gives the conclusions.

\section{Three-meson doorway mechanisms}
\label{TMDWM}

\begin{figure}
\begin{center}
\mbox{
\begin{picture}(210,110)(10,0) 
\Text(10,100)[b]{{\bf (a)}}
\Line(35,58)(60,58) \Line(35,62)(60,62) 
\Vertex(60,60){4}
\ArrowLine(60,60)(120,20) \Vertex(120,20){2}
\ArrowLine(60,60)(120,100) \Vertex(120,100){2}
\ArrowLine(60,60)(120,60) \Vertex(120,60){2}
\Text(120,107)[b]{$c$} \Text(120,15)[t]{$b$}
\Text(60,70)[b]{$a$} \Text(125,62)[l]{$d$}
\ArrowLine(120,60)(120,100) 
\ArrowLine(120,60)(120,20) 
\ArrowLine(120,100)(145,100) 
\ArrowLine(120,20)(145,20)
\Text(30,60)[r]{$p_a$}   
\Text(150,100)[l]{$p_c$} 
\Text(150,20)[l]{$p_b$}  
\Text(92,88)[br]{$p_y$} \Text(100,65)[b]{$p_z$}
\Text(95,34)[tr]{$p_x$}
\Text(125,84)[l]{$p_w$} \Text(125,40)[l]{$p_v$}
\end{picture}
\begin{picture}(210,100)(10,0) 
\Text(10,100)[b]{{\bf (b)}}
\Line(35,58)(60,58) \Line(35,62)(60,62) 
\Vertex(60,60){4}
\ArrowLine(60,60)(120,20) \Vertex(120,20){2}
\ArrowLine(60,60)(120,100) \Vertex(120,100){2}
\ArrowLine(60,60)(120,60) \Vertex(120,60){2}
\Text(120,107)[b]{$c$} \Text(120,15)[t]{$b$}
\Text(60,70)[b]{$a$} \Text(125,62)[l]{$d$}
\ArrowLine(120,60)(120,100) 
\ArrowLine(120,60)(120,20) 
\ArrowLine(120,100)(145,100) 
\ArrowLine(120,20)(145,20) 
\SetWidth{0.8} \Line(102,57)(108,63) \Line(108,57)(102,63)
\Line(106,94)(104,86) \Line(109,89)(101,91)
\Line(106,26)(104,34) \Line(109,31)(101,29)
\Text(30,60)[r]{$(m_a,{\bf 0})$}
\Text(150,100)[l]{$(E_c,-{\bf k_b})$} 
\Text(150,20)[l]{$(E_b,{\bf k_b})$}
\Text(92,88)[br]{$(E_y,{\bf k_y})$} 
\Text(100,65)[b]{$(E_z,{\bf k_z})$}
\Text(95,34)[tr]{$(E_x,{\bf k_x})$}
\Text(125,84)[l]{$p_w$} \Text(125,40)[l]{$p_v$}
\end{picture}
}
\end{center}
\vspace*{-5mm}
\caption{\label{FigTMDWM}
(a) The generic three-meson-doorway diagram;
(b) unitarity approximation. 
The notation $p_n=(E_n,{\bf k_n})$ is used  
for the four--momenta of the particles $n=a,b,c,x,y,z,v,w$ in the CMS. 
}
\end{figure}
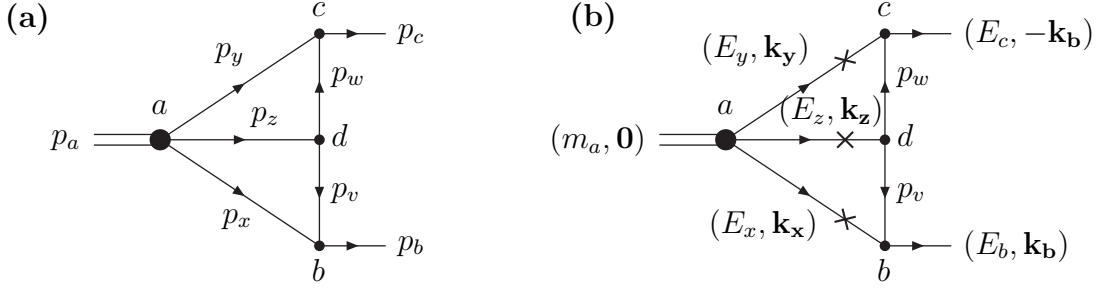

  The generic three-meson-doorway diagram is shown in Fig.~\ref{FigTMDWM}
which also defines the notation in terms of four and three vectors.  The
crosses in Fig.~\ref{FigTMDWM}(b) denote the unitarity approximation for
which all the three $s$-channel particles are on their mass shell.  The
available data for the annihilation into three mesons 
relevant for the first step of the three-meson-doorway mechanisms 
are summarized in Table~\ref{TabCoup3} of Appendix~\ref{App3MA}.

The basic expression for the amplitude without spin is
\begin{equation}
T = \frac {i^2}{(2 \pi)^8} \int \!\!\!\int \!
 \frac { g_a g_b g_c g_d \, d^4p_x\,d^4p_y}
 {(p_x^2-m_x^2+i\epsilon)(p_y^2-m_y^2+i\epsilon)(p_z^2-m_z^2+i\epsilon)
  (p_v^2-m_v^2+i\epsilon)(p_w^2-m_w^2+i\epsilon)}.
\label{T3MIS}
\end{equation}
where $g_a$, $g_b$, $g_c$, and $g_d$ are the coupling constants corresponding
to the vertices $a,b,c,d$ in Fig.~\ref{FigTMDWM}(a).
In the on-shell or unitarity approximation Fig.~\ref{FigTMDWM}(b),
the amplitude $T_{UA}$ has the form
\begin{eqnarray}
  T_{UA} & = &
  \frac{i g_a g_b g_c g_d}{2(2\pi)^5} \;
  \int
  \frac{d\Phi_3(p_a,p_x,p_y,p_z)}
       {(p_v^2-m_v^2+i\epsilon)(p_w^2-m_w^2+i\epsilon)}
  \label{T3UA}
\end{eqnarray}
where $d\Phi_3$ is the phase space of the intermediate three-meson state 
\begin{eqnarray}
  d\Phi_3(p_a,p_x,p_y,p_z) & = &
  \delta(p_a - p_x - p_y - p_z)
  \frac{d^3{\mathbf k}_x}{2E_x}
  \frac{d^3{\mathbf k}_y}{2E_y}
  \frac{d^3{\mathbf k}_z}{2E_z}
\label{PhSp}
\end{eqnarray}
The extension of Eqs.(\ref{T3MIS},\ref{T3UA}) to the case of particles with spin 
is straightforward.  The details for the vertex spin structure and coupling
constants used are given in Appendix~\ref{AppVert}. The spin formalism  
for the overall amplitude of Fig.\ref{FigTMDWM} is summarized in 
Appendix~\ref{AppSpinfact}.

  The leading two-loop mechanisms for the reactions 
$p\bar{p}\to\phi\pi^0$  correspond to the diagrams in Table \ref{TabSumphipi}.  
  The $\phi\pi$ channel has $I^G=1^+$ and originates from the  
$p\bar{p}(^3S_1)$ state $J^{PC}=1^{--}$.
Here we expect that the $\rho\pi\pi$ doorway mechanism is important 
because the $\rho\pi\pi$ system has the largest phase space among all three 
meson states with positive $G$-parity and is strongly produced in $p\bar{p}$ 
annihilation (see Appendix~\ref{AppVert}, Table~\ref{TabCoup3}).      
  The two-loop diagrams in Table \ref{TabSumphipi} all proceed through the same 
doorway mesons $(\rho\pi\pi)$ but differ by the mesons exchanged in the $t$-channel. 
The $\pi\pi$ subsystem in the intermediate state has total isospin $I=0$ 
and total angular momentum $J=0$. Thus there is no danger of double counting when 
this three-meson doorway mechanism is added to the two-meson $\rho\rho$ doorway 
mechanism which is known to be very important \cite{LMR}.

\begin{table}[bp] 
\begin{center}
\begin{tabular}{lr@{.}l}
\hline 
 mechanism \VS{8mm} & 
 \multicolumn{2}{c}
 {${\displaystyle | T / g_{a\to\phi\pi}^{\mathrm{tree}} | }$} 
\\[2mm]
\hline
\raisebox{-15mm}{\FigDiag{$\pi^{\mp}$}{$\rho^0$}{$\pi^{\pm}$}
                {$\ \rho^{\pm}$}{$\ \omega$}
                {$^3S_1$}{$\phi$}{$\pi^0$}}
 & 2&91   
\\
\hline
\raisebox{-15mm}{\FigDiag{$\rho^0$}{$\pi^{\pm}$}{$\pi^{\mp}$}
                {$\ \pi^0$}{$\ \rho^{\mp}$}
                {$^3S_1$}{$\phi$}{$\pi^0$}}
 & 1&35    
\\
\hline 
\raisebox{-15mm}{\FigDiagA{$\rho$}{$\rho$}{$\pi$}{$^3S_1$}{$\phi$}{$\pi$}} & 
   0&6 -- 2.6  
\\
\hline
\raisebox{-15mm}{\FigDiagA{$\bar{K}$}{$K^*$}{$K$}{$^3S_1$}{$\phi$}{$\pi$}} & 
   2&0 -- 3.3  
\\
\hline
 \VS{8mm} Experiment  &  7&0 $\pm 0.4$ \cite{Ams98}  
\\[2mm]
\hline
\end{tabular}
\caption{\label{TabSumphipi} 
Two-loop and one-loop diagrams for the reaction $p\bar{p}(^1S_0)\to\phi\pi$. 
The corresponding amplitudes $T$ are normalized to  
the tree-level amplitude $g_{a\to\phi\pi}^{\mathrm{tree}}$ 
from $\omega\phi$ mixing. 
}
\end{center}
\end{table}
 
\begin{table}[bp] 
\ \ \\
\VS{-30mm} 
\begin{center}
\hspace*{-10mm}
\begin{tabular}{lr@{.}l}
\hline 
 mechanism \VS{8mm} & 
 \multicolumn{2}{c}
 {${\displaystyle | T / g_{a\to\phi\rho}^{\mathrm{tree}} | }$} 
\\[2mm]
\hline
\raisebox{-15mm}{
\FigDiag{$\pi$}{$\pi$}{$\pi$}{$\rho$}{$\omega$}{$^1S_0$}{$\phi$}{$\rho$}} & 
   0&82    
\\
\hline
\raisebox{-15mm}{
\FigDiag{$\pi$}{$\pi$}{$\omega$}{$\rho$}{$\pi$}{$^1S_0$}{$\phi$}{$\rho$}} &
   0&74  
\\
\hline
\raisebox{-15mm}{\FigDiag{$\pi$}{$\omega$}{$\pi$}{$\rho$}{$\pi$}{$^1S_0$}{$\phi$}{$\rho$}} & 
   0&96  
\\
\hline
\raisebox{-15mm}{\FigDiag{$\pi$}{$\pi$}{$\pi$}{$\rho$}{$\pi$}{$^1S_0$}{$\phi$}{$\rho$}} & 
   0&18  
\\
\hline 
\raisebox{-15mm}{\FigDiagA{$\rho$}{$\omega$}{$\pi$}{$^1S_0$}{$\phi$}{$\rho$}} & 
   0&6   
\\
\hline
  \VS{8mm} Experiment &  4&9 $\pm 0.8$ \cite{AST91}   
\\[2mm]
\hline
\end{tabular}
\caption{\label{TabSumphirho} 
Two-loop and one-loop diagrams for the reaction $p\bar{p}(^1S_0)\to\phi\rho$. 
The corresponding amplitudes $T$ are normalized to  
the tree-level amplitude $g_{a\to\phi\rho}^{\mathrm{tree}}$ 
from $\omega\phi$ mixing. 
}
\end{center}
\end{table}

 Spin effects have been calculated by introducing scalar invariant functions 
as described in Appendix~\ref{AppSpinfact}. 
As a cross check, helicity amplitudes for the full amplitude $T$ have been 
evaluated, squared and summed.    
Several charge configurations for the intermediate states, 
see Table \ref{TabSumphipi}, add coherently leading to an enhancement of  
the two-loop contributions.
  The corresponding isospin factors are collected in Table~\ref{TabIF} 
of Appendix~\ref{InvAmpl}.  
  Because the $t$-channel particles can reach the mass shell, 
the unitarity approximation acquires a real part.  
A similar situation has been encountered already in the one loop
calculation for the two-meson doorway mechanism \cite{LMR}.

  For the kinematical situations where the $t$-channel particles are off-shell
we have introduced monopole form factors
\begin{equation}
 F_b(\lambda_v) = \frac{\lambda_v^2-m_v^2}{\lambda_v^2-p_v^2} =
                  \frac{\lambda_v^2-m_v^2}{\lambda_v^2-(p_b-p_x)^2}
\label{ffvb} 
\end{equation} 
\begin{equation}
 F_c(\lambda_w) = \frac{\lambda_w^2-m_w^2}{\lambda_w^2-p_w^2} = 
                  \frac{\lambda_w^2-m_w^2}{\lambda_w^2-(p_b+p_y)^2}.
\label{ffvc} 
\end{equation}
   The parameters have been varied in the range
$\lambda_v,\lambda_w=(1.0-1.5)\;$GeV, similar to the one loop
calculation \cite{LMR}.  The form factors reduce the unitarity amplitude
by about a factor two.  For a calculation beyond the unitarity
approximation form factors for the $s$-channel doorway mesons must be
introduced as well.  
   The corresponding off--shell contributions are expected to be comparable 
to the unitarity amplitude, similar to the detailed evaluations done in 
the one-loop case \cite{LMR}. 

\newcommand{\YTthree}{
\mbox{ 
\begin{picture}(10,10)
\put(0,0){\line(1,0){15}}
\put(0,5){\line(1,0){15}}
\put(0,0){\line(1,0){5}}
\put(0,0){\line(0,1){5}}
\put(5,0){\line(0,1){5}}
\put(10,0){\line(0,1){5}}
\put(15,0){\line(0,1){5}}
\end{picture}
}
}

\newcommand{\YTtwoone}{
\raisebox{-0.5mm}{ 
\begin{picture}(10,10)(0,-5)
\put(0,0){\line(1,0){10}}
\put(0,5){\line(1,0){10}}
\put(0,-5){\line(1,0){5}}
\put(0,-5){\line(0,1){10}}
\put(5,-5){\line(0,1){10}}
\put(10,0){\line(0,1){5}}
\end{picture}
}
}

   Turning to the $\phi\rho$ channel which has $I^G=1^-$ and originates from the  
$p\bar{p}(^1S_0)$ state $J^{PC}=0^{-+}$, we expect that 
the $\pi\pi\pi$ and $\omega\pi\pi$ doorway mechanisms are important 
because these intermediate states have the largest phase space among 
all three meson states with negative $G$-parity.   
Several states for the $\pi\pi\pi$ system are possible in this case 
which can be classified by the symmetry of the isospin wave function. 
   The completely symmetric isospin wave function corresponding to the 
Young tableau \YTthree\ has isospin $I=1$ (another completely symmetric state $I=3$ 
is excluded by isospin conservation) \cite{Pais60}, with the space part being also 
completely symmetric. In Appendix~\ref{AppVert}, the notation $g^A_{3\pi}$ is used  
for the corresponding coupling constant.   
   The state with the mixed symmetry of the isospin part \YTtwoone can also   
have total isospin $I=1$, the corresponding coupling constant is $g^B_{3\pi}$ 
(see Appendix~\ref{AppVert} for details).  In this case the $3\pi$ state always 
contains a pion pair with total isospin $I_{\pi\pi}=1$ and odd relative angular 
momentum. It is likely that this configuration is saturated by the 
$\rho^{\pm}\pi^{\mp}$ channel  (the channel $\rho^0\pi^0$ is not coupled 
to the $\phi\rho^0$ because of wrong $C$-parity). 
We shall drop such three--meson intermediate states to avoid double counting. 
    The completely antisymmetric isospin wave function has 
total isospin $I=0$ and cannot occur in the annihilation into $\phi\pi$.   
  Concerning the intermediate state $\omega\pi\pi$, its isospin structure is 
completely determined by the total isospin of the final state $I=1$ which  
is equal to the isospin of the pion pair. Therefore one can expect that 
this intermediate state is saturated by the one-loop intermediate state 
$\omega\rho^0$.   
For the purpose of information the tables show all relevant amplitudes 
calculated separately.  The two-loop contributions are fairly sizable. 

\section{Results}
\label{Res}
 
  The complete results for the $\phi\pi^0$ and $\phi\rho^0$ channels are 
summarized in Tables~\ref{TabSumphipi}, \ref{TabSumphirho}, and \ref{Resphi}.  
The three-meson doorway states shown are the ones leading to the 
biggest contributions.  
  The contributions from other intermediate states have been calculated and were 
found to be negligible \cite{SvR}.  
  The 'tree-level' amplitudes correspond to the $\omega\phi$ mixing which is 
proportional to the deviation of the physical mixing angle 
$\Theta=37.6^o$ \cite{PDG} from the the ideal one $\Theta_i=35.3^o$: 
\begin{equation}
    g^{\mathrm{tree}}_{a\to\phi X} = 
    \tan{(\Theta-\Theta_i)} \cdot g_{a\to\omega X} \quad . 
\end{equation}

  The on-shell values of the vertex functions occurring in the calculation
are constrained directly by experimental information.  For the annihilation
vertices, the data are shown in Table~\ref{TabCoup3} and the
corresponding coupling constants have been parametrized as described in
Appendix~\ref{App3MA}.  The remaining vertices have been calculated from the
measured decay widths of the corresponding mesons and are collected in 
Appendix~\ref{App2MD}. 
The values shown in Table~\ref{Resphi} do not include 
the form factors of Eqs.(\ref{ffvb},\ref{ffvc}) which lead  
to a reduction by about a factor of two.    
   This should be quite a reliable approximation since this   
reduction is expected to be partially compensated by
contributions originating from off-shell $s$-channel propagation,  
similarly to the case of two-meson doorway mechanism \cite{LMR}, 
as mentioned in the context of Eqs.(\ref{ffvb},\ref{ffvc}).  

  In general, major cancellations between amplitudes corresponding to different 
intermediate states are not likely to occur for the unitarity approximation.  
Because the main contribution in this case comes from the absorptive part of 
loop diagrams, the situation is very different from the well known 
case of Lipkin cancellations \cite{Lip84} where, contrary to our situation,  
threshold effects are negligible and specific intermediate states interfere 
distructively.            
In the case of the $\phi\pi$ channel the situation has been discussed 
in detail \cite{LMR} for the one-loop mechanism.  

   For the case of the $\rho\rho$ doorway contribution to $\phi\pi^0$ we have 
evaluated the full range of the possible coupling constants, see
Appendix~\ref{pprhorho}, complementing the results in \cite{SvR}. 
The corresponding range is indicated in Table~\ref{Resphi}.

\begin{table}
\begin{center}
\begin{tabular}{l|l}
\hline
Reaction / Mechanism             
      &  $BR \cdot 10^4$  
\\  
\hline
$\pp\to\phi\pi^0$ (tree-level)   
      &  0.13            
\\
$\pp\to K^*\bar{K}\to\phi\pi^0$  
      &  $0.5 - 1.4$             
\\
$\pp\to\rho\rho\to\phi\pi^0$     
      &  $0.05 - 2.0$                 
\\
$\pp\to\phi\pi^0$ (1-loop)   
      &  $0.9 - 5.1$             
\\
$\pp\to\rho^0\pi\pi\to\phi\pi^0$ 
      &  $0.22$                  
\\
$\pp\to\pi\rho^0\pi\to\phi\pi^0$ 
      &  1.1
\\
$\pp\to\phi\pi^0$ (1-loop and 2-loop)   
      &  $5.9 - 17$               
\\
$\pp\to\phi\pi^0$ ({\bf experiment})  
      &  $6.5\pm 0.7$ \quad \cite{Ams98}    
\\ 
      &  $7.6\pm 0.6$ \quad \cite{OBEL98a} 
\\
      &  $4.0\pm 0.8$ \quad \cite{AST91}
\\  
\hline
$\pp\to\phi\rho^0$ (tree-level)   
      & 0.14
\\
$\pp\to\phi\rho^0$ (one-loop)     
      & 0.1   
\\
$\pp\to\omega\pi\pi\to\phi\rho^0$ 
      & 0.08                      
\\
$\pp\to\pi\pi\pi\to\phi\rho^0$    
      & 0.10
\\
$\pp\to\phi\rho^0$ (1-loop and 2-loop)   
      & $\sim 2$                  
\\ 
$\pp\to\phi\rho^0$ ({\bf experiment})  
      & $3.4\pm 1.0$ \quad \cite{AST91} 
\\  
\hline
$\pp\to\phi\eta$ (tree-level)    
      & 0.2 
\\
$\pp\to K^*\bar{K}\to\phi\eta$   
      & $1.0$                    
\\
$\pp\to\phi\eta$ ({\bf experiment})  
      & $0.78\pm 0.21$  \quad \cite{CBC95}   
\\  
\hline
\end{tabular}
\end{center}
\caption{The branching ratios $BR$ calculated for various doorway 
mechanisms in comparison with the experimental data (in units $10^{-4}$). 
The values marked 1-loop and 2-loop contributions correspond to 
different ways of adding the amplitudes coherently.   
}
\label{Resphi}
\end{table}

   For the OZI--violating final state $\phi\pi^0$  
the calculated branching ratio in Table~\ref{Resphi} is well within the
experimental range.  The three--meson doorway contributions are comparable with 
the two--meson ones.    
  The range of theoretical predictions when adding the amplitudes coherently 
now easily includes the experimental branching ratios while the one-loop results 
alone are somewhat low. 
  For the $\phi\eta$ channel, the two-meson-doorway mechanism with 
the $K^*\bar{K}$ intermediate state has been found to be comparable with 
the experimental data.  
  The two-loop calculations for the $\phi\rho$ final states are also reported in 
Table~\ref{Resphi}.  As has been mentioned at the end of Section~\ref{TMDWM}, 
two of the diagrams involve double counting with one-loop mechanisms, which is not 
easily quantified.  However, the two-loop contributions obviously improve 
the comparison with the experimental branching ratio.  

  At the end of this section we would like to mention that we have also
evaluated two and three-meson doorway contributions for a number of
two meson final states ($\pi\pi$, $\rho\pi$, $\rho\rho$ and
$\rho\omega$) without hidden strangeness.  
It is gratifying to observe that all the doorway contributions calculated have 
turned out to be relatively small when compared to the experimental rates for the
corresponding annihilation channels.

\section{Conclusions}
\label{Concl}

   We have found that the observed OZI violating enhancement of $\phi$ meson
production at rest can be naturally explained by two and three meson
doorway contributions.  
   In our analysis, there appears to be no need to introduce a large
$s\bar{s}$ fraction into the nucleon wave function.  The doorway
calculations presented here are well constrained by experimental
information.
   In the first step of the doorway mechanism the annihilation rates into
non-strange mesons enter. For annihilation at rest these transition rates are 
well measured.
This is particularly true for the largest observed OZI violation in 
$p\bar{p}\to\phi\pi^0$ where detailed information on the spin--isospin 
dependence of the amplitudes for the annihilation $p\bar{p}\to K^*\bar{K}$ 
exists.  Similarly the meson decay vertices occuring in the second step 
of the one--loop doorway mechanism are directly constrained by the measured 
decay widths. The leading one--loop contribution is thus well determined.  
In the present paper we have shown that the dominant three--meson doorway 
mechanisms (two loops) for $p\bar{p}\to\phi\pi$ are of similar size as the 
one loop contributions.  It is therefore established that the full calculation 
leaves ample space for accommodating any remaining descrepancy with the measured 
branching ratio.   
   At the same time two and three-meson doorway contributions to all the other 
channels involving $\phi$ mesons in the final state  are small but not negligible 
due to interference, which again is in agreement with measured branching ratios.
   Four-meson-doorway contributions and higher are expected to be
negligible due to progressively vanishing probability to rearrange  
non-strange multi-meson intermediate states into two--meson final states.  
We therefore believe that the
present multiple doorway analysis is qualitatively exhaustive for
nucleon-antinucleon annihilation into $\phi$ mesons at rest.
   Extending these calculations towards higher energies seems desirable.
However, the experimental information on the energy dependence of the
production of the intermediate states is far less detailed and large
uncertainties in the corresponding predictions appear to be unavoidable.

\section*{Acknowledgment}

We are grateful to M.~Sapozhnikov for stimulating discussions.
This paper was supported in part by the Swiss National Science Foundation. 


\appendix 

\renewcommand{\theequation}{A\arabic{equation}}
\setcounter{equation}{0}
\section{Coupling constants}
\label{AppVert} 

  In this appendix, we collect the coupling constants for $\bar{p}p$
annihilation at rest into various two and three-meson annihilation channels
and the required coupling constants for meson decays.  For most cases, 
the couplings can be expressed in terms of partial decay widths which are known 
from experiment and provide a model-independent input to the calculations of the
doorway mechanisms.

\subsection{Vertices for two-particle decays} 
\label{App2MD}

  The amplitudes corresponding to the transitions $a\to b+c$
involving pseudoscalar fields $\phi$ and vector fields $V^\mu$
with minimal number of derivatives have the following form in momentum space
\begin{eqnarray}
\langle\phi_b\phi_c|T|V_a\rangle
   & = & g_{V\phi\phi}\;\ve_a\cdot(p_b-p_c) \label{coup21}
\\
\langle V_b\phi_c|T|V_a\rangle
   & = & g_{VV\phi}\;\epsilon_{\mu\nu\alpha\beta} \;
                p_a^\mu\ve_a^\nu p_b^\alpha\ve_b^\beta \label{coup23}
\\
\langle V_bV_c|T|V_a\rangle
   & = &   g_V^{(1)}\;p_a\cdot\ve_b\;\ve_a\cdot\ve_c
          +g_V^{(2)} \; p_a\cdot\ve_c \; \ve_a\cdot\ve_b
          +g_V^{(3)} \; p_b\cdot\ve_a \; \ve_b\cdot\ve_c \label{coup24}
\end{eqnarray}
where $p_a$, $p_b$, $p_c$ are the corresponding four-momenta,
$\ve_a$, $\ve_b$, $\ve_c$ are the polarization vectors and
$\epsilon^{\mu\nu\alpha\beta}$ is the totally antisymmetric
Levi-Civit\`a tensor.
  The decay widths $\Gamma_{a\to b+c}$ are related to the corresponding
coupling constants $g_{abs}$ by
\begin{eqnarray}
  \Gamma_{a\to b+c} & = & \frac{g_{abc}^2 f(k_{bc}) k_{bc}}{8\pi m_a^2}
  \label{Gabc}
\\
  k_{bc} & = & \frac{\sqrt{(m_a^2-(m_b+m_c)^2)(m_a^2-(m_b-m_c)^2)}}{2m_a}
  \label{kbc}
\end{eqnarray}
where $k_{bc}$ is the CMS momentum of the particles $b$ ans $c$,
$m_a$, $m_b$, and $m_c$ are the corresponding particle masses and
$f(k_{bc})$ are the spin-weight functions defined in Table~\ref{TabSF2}. 

\begin{table}[hbtp]
\begin{center}
\begin{tabular}{|l|c|}
\hline
  Reaction $a\to b+c$ &    Spin-weight functions $f(k_{bc})$
\\[2mm]
\hline
  $0^+\to 0^\pm 0^\pm$ & 1
\\[2mm] 
  $0^-\to 1^- 1^-$ & $2 m_a^2 k_{bc}^2$  
\\[2mm] 
  $1^-\to 0^\pm 0^\pm$ & $\frac{4}{3} k_{bc}^2$   
\\[2mm]
$0^\pm\to 1^- 0^\pm$ & ${\displaystyle \frac{4 m_a^2 k_{bc}^2}{3 m_b^2}}$ 
\\[2mm]  
$1^-\to 1^- 0^-$     & $\frac{2}{3} m_a^2 k_{bc}^2$ 
\\[2mm] 
$1^-\to 1^- 1^-$ (a)
      & ${\displaystyle \frac{m_a^2(m_a^2+2m_b^2+2m_c^2)k_{bc}^2}{3m_b^2m_c^2}}$ 
\\[2mm] 
$1^-\to 1^- 1^-$ (b)
      & ${\displaystyle \frac{2k_{bc}^2}{3} 
                        \left(3+\frac{m_a^2k_{bc}^2}{m_b^2m_c^2}\right)}$ 
\\[2mm]  
\hline
\end{tabular}
\end{center}
\caption{\small
The spin-weight functions $f(k_{bc})$, Eq.(\protect{\ref{Gabc}}),
for two particle decays. 
For the three-vector-meson vertex Eq.(\protect\ref{coup24}), the two cases
correspond to the situations considered in Section~\protect\ref{pprhorho}:
(a) $g_v^{(12)}=g_v^{(1)}=-g_v^{(2)}$, $g_v^{(3)}=0$ and 
(b) $g_v^{(1)}=g_v^{(2)}=0$, $g_v^{(3)}\neq 0$, both assuming $m_b=m_c$.
}
\label{TabSF2}
\end{table}

\begin{table}[hbtp]
\begin{center} 

\begin{tabular}{|l|cl|l|}
\hline 
 Process  &  Ref.  &  Branching ratio $BR$  &  $g_{abc}$ \\   
\hline 
 $p\bar{p}(^1\!S_0) \to \phi \rho$  & 
  \cite{AST91}  &  $(3.4 \pm 1.0) \cdot 10^{-4}$  & 
            {$g_{a\to\phi\rho} = 8.71 \cdot 10^{-4} $}  
\\ 
\hline 
 $p\bar{p}(liq.) \to \omega\rho^0_{\to\pi+\pi-}$  & 
  \cite{Biz69}     &  $(2.26 \pm 0.23) \cdot 10^{-2}$  & 
            {$g_{a\to\omega\rho} = 2.77 \cdot 10^{-3} $ }
\\ 
\hline 
 $p\bar{p}(S\to ^3\!S_1) \to \phi \pi^0$  &                          
  \cite{Ams98}     &  $(6.5 \pm 0.6) \cdot 10^{-4}$  &               
            {$g_{a\to\phi\pi} = 3.43 \cdot 10^{-4} $ }
\\ 
 & \cite{CBC95}     &  $(5.5 \pm 0.7) \cdot 10^{-4}$  & 
\\ 
 & \cite{OBEL98a}  &  $(7.57 \pm 0.62) \cdot 10^{-4}$  & 
\\ 
 & \cite{AST91} &  $(4.0 \pm 0.8) \cdot 10^{-4}$  & 
\\ 
\hline 
 $p\bar{p}(S\to ^3\!S_1) \to \omega \pi^0$  &                       
   \cite{CBC93}     &  $(5.7 \pm 0.5) \cdot 10^{-3}$  & 
            {$g_{a\to\omega\pi} = 7.59 \cdot 10^{-4} $ }
\\ 
 & \cite{Ch88}   &  $(5.2 \pm 0.5) \cdot 10^{-3}$  & 
\\ 
\hline 
 $\pp\to\rho\rho$ &
   \cite{2MDM0}  &  $2.4 \cdot 10^{-2}$   &    
            $g_{\rho\rho}^{(1)} = 8.21\cdot10^{-4}$ 
\\
  &    &   & 
            $g_{\rho\rho}^{(2)}=9.79\cdot10^{-4} \;\GeV^{-1}$  
\\
\hline 
 $\pp\to K^{*}\bar{K}, \bar{K}^*K$ & 
   \cite{Co67}  & $ 0.23 \cdot 10^{-2}$ &                         
   $g_{K^{*}\bar{K}} = 7.0 \cdot 10^{-4} \;\GeV^{-1}$   
\\ 
\hline            
\end{tabular}
\end{center}
\caption{\small
The experimental branching ratios for the two-meson $\pp$ annihilation  
at rest and the corresponding coupling constants used in the present calculations. 
The coupling constants are normalized to the total width of the 
atomic $(p\bar{p})_{1S}$ state $\Gamma_{(p\bar{p})_{1s}}=1\;$keV  
(the singlet--to--triplet ratio $1:3$ is assumed for the $p\bar{p}$ spin 
fractions).
}
\label{TabCoup2}
\end{table}

The following coupling constants for the meson decays were used in the 
present calculations: 
$g_{\rho\pi\pi}=6.00$,
$g_{\phi K\bar{K}}=4.6$, 
$g_{K^*K\pi}=5.54$,
$g_{\phi\rho\pi}=1.86\;\GeV^{-1}$. 
  The numerical values of the coupling constants for the two-meson 
$p\bar{p}$ annihilation are summarized in Tab.~\ref{TabCoup2} 
together with the corresponding experimental branching ratios.
For the sake of convenience, the $p\bar{p}$ annihilation coupling constants 
are normalized to the partial widths of the ground state of the $p\bar{p}$ atom. 
These partial widths are related to the corresponding annihilation 
cross sections $\sigma_{p\bar{p}\to b+c}$ by 
\begin{eqnarray}  
   \Gamma_{p\bar{p}\to b+c} & = & 
   (v \sigma_{p\bar{p}\to b+c})_{v\to 0} |\psi_{1S}(0)|^2 
\end{eqnarray}  
where $v$ is the relative velocity and 
$|\psi_{1S}(0)|^2=\frac{\alpha^3 m_p^3}{8\pi}$ is the probability density for 
the $1S$ atomic state at zero separation between $p$ and $\bar{p}$. This gives the 
following relation between the coupling constants listed in Table \ref{TabCoup2} 
and the $S$-wave annihilation amplitudes $g_{p+\bar{p}\to b+c}$ at zero energy:   
\begin{eqnarray}  
   g_{a\to b+c}^2 & = & g_{p+\bar{p}\to b+c}^2 \frac{|\psi_{1S}(0)|^2}{m_p} 
                    =   g_{p+\bar{p}\to b+c}^2 \frac{\alpha^3 m_p^2}{8\pi}  
\end{eqnarray}

\subsection{Vertices for three-meson annihilation} 
\label{App3MA}

In the case of three-particle transitions $a\to x+y+z$ we consider
reactions of the following types
\begin{eqnarray} 
       0^- & \to & 0^- \, 0^- \, 0^- \nonumber \\ 
       0^- & \to & 1^- \, 0^- \, 0^- \nonumber \\ 
       1^- & \to & 1^- \, 0^- \, 0^- \nonumber    
\end{eqnarray}
The corresponding amplitudes with a minimal number of derivatives read 
\begin{eqnarray}
   \langle\phi_x\phi_y\phi_z|T|\phi_a\rangle & = & g_{pppp}
\label{gPPPP} \\
   \langle\phi_x\phi_yV_z|T|\phi_a\rangle    & = & g_{pppv}\;
   \epsilon_{\mu\nu\alpha\beta}\;p_a^\mu p_x^\nu p_y^\alpha\ve_z^\beta
\label{gPVPP} \\
   \langle\phi_x\phi_yV_z|T|V_a\rangle  & = & g_{vppv} \; \ve_a \cdot \ve_z
\label{gVVPP} 
\end{eqnarray}
The coupling constants $g_{axyz}$ are related to the three-body transition
widths $\Gamma_{a\to xyz}$ by
\begin{equation}
   \Gamma_{a\to xyz} = \frac{g_{axyz}^2}{2m_a (2\pi)^5}
                       \int w \; d\Phi_3.
\end{equation}
where $w$ are the kinematical factors given in Table~\ref{TabSF3} and
the three-body phase space $d\Phi_3$ is defined by Eq.(\ref{PhSp}).  

  For the $p\bar{p}(0^-)\to\pi\pi\pi$ vertex we have two cases considered in 
Sect.~\ref{TMDWM}.  The vertex that is completely symmetric in the isospin of the 
$\pi\pi\pi$ system corresponds to Eq.(\ref{gPPPP}) with the coupling constant 
$g_{3\pi}^A$.  The vertex with the mixed symmetry is given by 
\begin{eqnarray}  
   \langle \pi^+\pi^0\pi^-|T|p\bar{p}(J=0^-,I=1)\rangle & = & 
   g_{3\pi}^B ( p_0\cdot p_+ + p_0\cdot p_- -2 p_+\cdot p_+) 
\label{BgPPPP}       
\end{eqnarray}  
where $p_+$, $p_0$, $p_-$ are the four-momenta of the corresponding pions. 

  The numerical values of the coupling constants for the three-meson 
$p\bar{p}$ annihilation are summarized in Tab.~\ref{TabCoup3} 
together with the corresponding experimental branching ratios.

\begin{table}[hbtp] 
\begin{center}
\begin{tabular}{|l|c|} \hline
Reaction        &       Spin-weight factor 
\\[2mm]
$a\to x+y+z$            & $w$ 
\\[2mm]
\hline
$0^- \to 0^- \, 0^- \, 0^-$ & 1 
\\[2mm]
$0^- \to 1^- \, 0^- \, 0^-$ & 
    $m_a^2 ({\mathbf k}_x^2 {\mathbf k}_y^2 - ({\mathbf k}_x{\mathbf k}_y)^2)$ 
\\[2mm]
$1^- \to 1^- \, 0^- \, 0^-$ &
                       ${\displaystyle 1+\frac13\frac{k_x^2}{m_x^2}}$ 
\\[2mm] 
\hline
\end{tabular}
\end{center}
\caption{
Spin--weight functions $w$ for different three-body final states. 
The ${\mathbf k}_x$ and ${\mathbf k}_y$ are the 3-momenta of particles 
$x$ and $y$ in the CMS.  
The interaction terms are defined in
Eqs.(\protect\ref{gPPPP}-\protect\ref{gVVPP}).
}
\label{TabSF3}
\end{table}

\begin{table}
\begin{center}
\begin{tabular}{|l|cl|l|}
\hline 
 Process  &  Ref.  &  Branching ratio $BR$  &  $|g_{abcd}|$ \\   
\hline 
 $p\bar{p}(liq.) \to \omega \pi\pi $  & 
  \cite{Biz69}  &  $0.066 \pm 0.006$  &    
\\ 
 $p\bar{p}(S\to \mbox{$^1\!S_0$}) \to \omega \rho^0_{\to\pi+\pi-} $  & 
  \cite{Biz69}  &  $0.0226 \pm 0.0023 $  & 
            $g_{p\bar{p}(^1S_0)\to\omega\pi\pi} = 0.21\;\mbox{\rm GeV}^{-3}$  
\\ 
\hline 
 $p\bar{p}(S) \to \pi^0\pi^+\pi^- $  & 
 \cite{May90} &  $0.066 \pm 0.008 $  &                 
\\ 
 $p\bar{p}(S\to\,^1\!S_0) \to (\pi^0\pi^+\pi^-)_{\mathrm ph.sp.} $  & 
  &  $0.066 \cdot (0.083 \pm 0.029) $  &    
            $g^A_{p\bar{p}(^1S_0)\to 3\pi} = 0.015 $  
\\ 
 $p\bar{p}(S\to\,^1\!S_0) \to \rho^{\pm}\pi^{\mp} $  & 
  &  $0.066 \cdot (0.014 \pm 0.006) $  &    
            $g^B_{p\bar{p}(^1S_0)\to(3\pi)} = 0.0056 $ 
\\ 
\hline 
 $p\bar{p}(liq.) \to \pi^0\pi^+\pi^- $  & 
   \cite{Fo68} &  $0.069 \pm 0.004 $  &                 
\\ 
\hline 
 $p\bar{p}(liq.) \to \pi^0\pi^0\pi^0 $  &  
  \cite{LEAR6} &  $(6.2 \pm 1.0)\cdot 10^{-3} $  &     
\\              
 $p\bar{p}(S\to\,^1\!S_0) \to \pi^0\pi^0\pi^0 $  & 
               &  $6.2\cdot 10^{-3} \cdot 0.54 $  &  
           $g^A_{p\bar{p}(^1S_0)\to 3\pi} = 0.010$   
\\              
\hline 
 $p\bar{p}(S\to\,^3\!S_1) \to \rho_{\to\pi^+\pi^-}\sigma_{\to\pi^+\pi^-} $  & 
  \cite{Ber97} &  $  7.61 \cdot 10^{-2} \cdot 0.50 $  &  
           $ g_{p\bar{p}(^3S_1)\to\rho\pi\pi}  = 0.043 $   
\\ 
\hline 
\end{tabular} 
\end{center}
\caption{\label{TabCoup3}
The experimental branching ratios for the three--meson $\pp$ annihilation  
at rest and the corresponding coupling constants used in the present calculations. 
The coupling constants are normalized to the total width of the 
atomic $(p\bar{p})_{1S}$ state $\Gamma_{(p\bar{p})_{1s}}=1\;$keV 
and the singlet--to--triplet ratio $1:3$ is assumed for the $p\bar{p}$ spin 
fractions.
}
\end{table}

\renewcommand{\theequation}{B\arabic{equation}}
\setcounter{equation}{0}
\section{Evaluation of Amplitudes with Spin}
\label{AppSpinfact}

The on-shell approximation Fig.~\ref{FigTMDWM}(b) for particles with spin leads to 
the following expression for the covariant amplitudes $T_{UA}$ 
replacing the spinless case of Eq.(\ref{T3UA})  
\begin{eqnarray}
  T_{UA} & = &
  \frac{i g_a g_b g_c g_d}{2(2\pi)^5} \;
  \int
  \frac{ w(p_x,p_y,p_z) d\Phi_3(p_a,p_x,p_y,p_z)}
       {(p_v^2-m_v^2+i\epsilon)(p_w^2-m_w^2+i\epsilon)}
  \label{T3UAspin}
\end{eqnarray} 
The evaluation of the spin--weight functions $w(p_x,p_y,p_z)$ can be done on 
the amplitude level using a decomposition of the spin functions into covariant 
tensor structures built of the external momenta $p_a$ and $p_b$.
The corresponding computations have been done using symbolic codes written
in MAPLE and were verified by hand for several cases. As a further, independent
check we have determined helicity amplitudes in an explicit
polarization basis, see e.g. Ref.~\cite{Gre84}, and calculated the
sum of the squared helicity amplitudes.

\subsection{Invariant Amplitudes and General Tensor Decomposition}
\label{InvAmpl}

The covariant integrals over
internal momenta $ p_x^\mu$ and $ p_y^\mu$ can generally be expressed in
terms of linearly independent tensors constructed of the external momenta,
$ p_a^\mu$ and $ p_b^\mu$ multiplied by invariant amplitudes.
For the one-loop diagrams (see Tables \ref{TabSumphipi} and \ref{TabSumphirho}) 
we introduce the following notation
\begin{equation}
 \langle  p_x^\mu\rangle = \int\
 f( p_x, p_a, p_b)\, p_x^\mu\,d^4\! p_x  \label{wint}
\end{equation}
where the scalar function $f( p_x, p_a, p_b)$ is defined by a direct 
calculation of the one loop diagram.   
The general form of this integral is given by
\begin{equation}
 \langle p_x^\mu\rangle=I_1^{(1)}\, p_a^\mu+I_2^{(1)}\, p_b^\mu \label{ans12}
\end{equation}
where the coefficients $I_1^{(1)}$ and $I_2^{(1)}$ can be found 
straightforwardly:
\begin{eqnarray}
 I_1^{(1)} & = & \langle \frac {k_bE_x-E_bk_xz_x}{m_ak_b} \rangle \label{I11} \\
 I_2^{(1)} & = & \langle \frac {k_xz_x}{k_b} \rangle \quad .      \label{I12}  
\end{eqnarray}
Here the notation $\langle\ldots\rangle$ is defined similarly to Eq.(\ref{wint}) 
with the same scalar function $f( p_x, p_a, p_b)$. 
The second order expressions in the one--loop case contain in addition two 
products of $ p_a^\mu$ and $ p_b^\mu$ and the metric tensor $g^{\mu\nu}$:  
\begin{equation}
\langle p_x^\mu p_x^\nu\rangle
=I_1^{(2)}\,g^{\mu\nu}+I_2^{(2)}\, p_a^\mu p_a^\nu
+I_3^{(2)}( p_a^\mu p_b^\nu+ p_b^\mu p_a^\nu)+I_4^{(2)}\, p_b^\mu p_b^\nu
\label{ans22}
\end{equation}
where 
\begin{eqnarray}
I_1^{(2)} & = & \langle -\frac {(1-z_x^2)k_x^2}2 \rangle \label{I21} 
\\
I_2^{(2)} & = & \langle \frac {2\,(k_x^2z_x^2+E_x^2)k_b^2-4\,E_bk_bE_xk_xz_x
                -(1-3\,z_x^2)m_b^2k_x^2}{2\,m_a^2k_b^2} \rangle 
\\
I_3^{(2)} & = & \langle \frac {((1-3\,z_x^2)E_bk_x+2\,k_bE_xz_x)k_x}{2\,m_ak_b^2} \rangle 
\\
I_4^{(2)} & = & \langle -\frac {(1-3\,z_x^2)k_x^2}{2\,k_b^2} \rangle \label{I24}.
\end{eqnarray}

\begin{figure}[hbtp]
\begin{center}
\begin{picture}(100,80)(-5,10) 
 \Text(10,60)[b]{(A)}
 \Line(25,30)(50,10) 
 \Line(25,30)(50,30) 
 \Line(50,50)(65,50)
 \SetWidth{1.8} 
 \Line(50,30)(50,10) 
 \Line(10,30)(25,30)
 \Line(50,50)(50,30) 
 \Line(25,30)(50,50) 
 \Line(50,10)(65,10)
 \Vertex(25,30){2}  
 \Vertex(50,50){2}  
 \Vertex(50,10){2}  
 \Vertex(50,30){2}
 \Text(6,31)[r]{$a$}  \Text(70,11)[l]{$b$}  \Text(70,51)[l]{$c$}
\end{picture}
\begin{picture}(100,80)(-5,10)
 \Text(10,60)[b]{(B)}
 \Line(10,30)(25,30) 
 \Line(25,30)(50,10) 
 \Line(25,30)(50,30) 
 \Line(50,50)(50,30)
 \SetWidth{1.8}
 \Line(25,30)(50,50) 
 \Line(50,30)(50,10) 
 \Line(50,10)(65,10) 
 \Line(50,50)(65,50)
 \Vertex(25,30){2}  
 \Vertex(50,50){2}  
 \Vertex(50,10){2}  
 \Vertex(50,30){2}
 \Text(6,31)[r]{$a$}  \Text(70,11)[l]{$b$}  \Text(70,51)[l]{$c$} 
\end{picture}
\\
\begin{picture}(100,80)(-5,10)
 \Text(10,60)[b]{(C)}
 \Line(10,30)(25,30) 
 \Line(25,30)(50,10) 
 \Line(25,30)(50,50) 
 \Line(25,30)(50,30)
 \SetWidth{1.8} 
 \Line(50,30)(50,10) 
 \Line(50,50)(50,30) 
 \Line(50,10)(65,10) 
 \Line(50,50)(65,50)
 \Vertex(25,30){2}  
 \Vertex(50,50){2}  
 \Vertex(50,10){2}  
 \Vertex(50,30){2}
 \Text(6,31)[r]{$a$} \Text(70,11)[l]{$b$} \Text(70,51)[l]{$c$} 
\end{picture}
\begin{picture}(100,80)(-5,10)
 \Text(10,60)[b]{(D)}
 \Line(50,30)(50,10) 
 \Line(25,30)(50,50) 
 \Line(25,30)(50,30) 
 \Line(50,50)(65,50)
 \SetWidth{1.8}
 \Line(10,30)(25,30) 
 \Line(50,50)(50,30) 
 \Line(25,30)(50,10) 
 \Line(50,10)(65,10)
 \Vertex(25,30){2}  
 \Vertex(50,50){2}  
 \Vertex(50,10){2}  
 \Vertex(50,30){2}
 \Text(6,31)[r]{$a$}  \Text(70,11)[l]{$b$}  \Text(70,51)[l]{$c$} 
\end{picture}
\end{center}
\caption{Different three-meson doorway processes involving 
pseudoscalar particles (thin lines) and vector particles (thick lines).}
\label{FigSWF}
\end{figure}
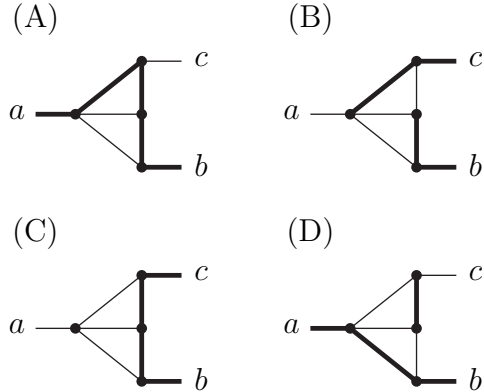

In the case of two-loop amplitudes we have more internal degrees of freedom, 
and Eq.(\ref{ans22}) is replaced by 
\begin{equation}
\langle  p_x^\mu p_y^\nu\rangle=\int\
f( p_x, p_y, p_a, p_b)\, p_x^\mu\, p_y^\nu\,d^4\! p_x\,d^4\! p_y.
\end{equation}
The tensor decomposition for this expression takes the following form
\begin{equation}
\langle p_x^\mu p_y^\nu\rangle
=I_1^{(xy)}\,g^{\mu\nu}+I_2^{(xy)}\, p_a^\mu p_a^\nu
+I_3^{(xy)}( p_a^\mu p_b^\nu+ p_b^\mu p_a^\nu)
+I_4^{(xy)}\, p_b^\mu p_b^\nu.
\label{ansxy}
\end{equation}
where
\begin{eqnarray}
I_1^{(xy)} & = & \langle -\frac {(Z-z_xz_y)k_xk_y}2 \label{Ixy1} \rangle 
\\
I_2^{(xy)} & = & \langle \frac {2\,(k_xk_yz_xz_y+E_xE_y)k_b^2-4\,E_bk_bE_xk_xz_x
                -(Z-3\,z_xz_y)m_b^2k_xk_y}{2\,m_a^2k_b^2} \rangle 
\\
I_3^{(xy)} & = & \langle \frac {((Z-3\,z_xz_y)E_bk_x+2\,k_bE_xz_y)k_y}{2\,m_ak_b^2} \rangle 
\\
I_4^{(xy)} & = & \langle -\frac {(Z-3\,z_xz_y)k_xk_y}{2\,k_b^2}\rangle . \label{Ixy4}
\end{eqnarray}
A generalisation to tensors of higher rank is straightforward. 
In particular, the following expressions for a rank three tensor appear 
in the two--loop diagrams:  
\begin{eqnarray}
I_7^{(yxx)} & = & \langle \frac{(z_xZ-z_y)k_x^2k_y}2
        -\frac{((Z-z_xz_y)E_xk_xk_y-(1-z_x^2)E_yk_x^2)k_b}{2E_b} \rangle 
\label{Iyxx7} \\
I_7^{(yxy)} & = & \langle \frac{(z_x-z_yZ)k_xk_y^2}2
        +\frac{((Z-z_xz_y)E_xk_xk_y-(1-z_y^2)E_xk_y^2)k_b}{2E_b} \rangle 
\label{Iyxy7}
\end{eqnarray}
The different topologies of the two-loop diagrams involving vector and 
pseudoscalar mesons are shown in Fig.\ref{FigSWF}.  
The corresponding complete spin weight functions Eq.(\ref{T3UAspin}) 
for the diagrams A--D are listed below. 

%
\begin{eqnarray}
& & \hspace*{-10mm} w_A(k_x,k_y,z_x,z_y)=(k_bE_y+E_ck_yz_y)(k_bE_x-E_bk_xz_x)+ \\
&+& \frac{1}{2m_ak_b}\left\{
   -\left[\left((z_x+z_yZ)E_ck_y+(z_xz_y+Z)k_bE_y\right)E_b^2
   +(z_xz_y+Z)E_c^2k_bE_x\right]k_xk_y+ \right.  \nonumber \\
&+& \left[\left((z_x+z_yZ)k_y^2+2m_y^2z_x\right)E_xk_x
    -\left((z_y+z_xZ)k_x^2+2m_x^2z_y\right)E_yk_y\right]k_b^2- \nonumber \\
&-& \left[\left((z_y^2+1)k_y^2+2m_y^2+(z_xz_y-Z)k_xk_y\right)E_x\right.+
    \nonumber \\
& & +\left.\left((z_x^2+1)k_x^2+2m_x^2+(z_xz_y-Z)k_xk_y\right)E_y\right]k_b^3+
    \nonumber \\
&+& \left[(z_y+z_xZ)E_c^2k_x^2k_y+\left((z_x+z_yZ)E_xk_xk_y^2
   -(z_y+z_xZ)E_yk_x^2k_y+ \right. \right. \nonumber \\
& & \left. +\left((2k_xz_xk_yz_y+(z_y^2+1)k_y^2)E_x
    +(2k_xz_xk_yz_y+(z_x^2+1)k_x^2)E_y\right)k_b\right)E_c+ \nonumber \\
& & +\left(2(k_xz_x+k_yz_y)E_xE_y+((z_y^2+1)k_y^2+2m_y^2
    +(z_xz_y-Z)k_yk_x)k_xz_x\right)k_b^2+ \nonumber \\
& & \left. +\left((z_xz_y+Z)E_xE_yk_xk_y+((z_xz_yZ-z_x^2-z_y^2-Z^2)k_y^2
    -(z_x^2+1)m_y^2)k_x^2\right)k_b\right]E_b- \nonumber \\
&-& \left[\left(2(k_xz_x+k_yz_y)E_xE_y+((z_x^2+1)k_x^2+2m_x^2
    +(z_xz_y-Z)k_xk_y)k_yz_y\right)k_b^2- \right. \nonumber \\
& & \left. \left. -\left((z_xz_y+Z)E_xE_yk_xk_y+((z_xz_yZ-z_x^2-z_y^2-Z^2)k_x^2
    -(z_y^2+1)m_x^2)k_y^2\right)k_b\right]E_c\right\}. \nonumber
\label{spin9}
\end{eqnarray}

\begin{eqnarray} 
w_B(k_x,k_y,z_x,z_y) & = & 
 -2\left(I_1^{(2)}+I_7^{(yxx)}\frac{E_b}{m_ak_b}\right)
  \left( p_a\cdot p_y\; p_c\cdot p_y- p_a\cdot p_c\;m_y^2\right)+ 
\nonumber  
\\ & &  \hspace*{-10mm}
  +2\left(I_1^{(xy)}+I_7^{(yxy)}\frac{E_b}{m_ak_b}\right)
  ( p_a\cdot p_y\; p_c\cdot p_x- p_a\cdot p_c\; p_x\cdot p_y)
\\
w_C(k_x,k_y,z_x,z_y) & = & 
  I_1^{(1)}( p_c\cdot p_w\; p_y\cdot p_v
  - p_c\cdot p_v\; p_y\cdot p_w)
  +I_1^{(2)} p_y\cdot p_w-I_1^{(xy)} p_y\cdot p_v+ 
\nonumber 
\\  & &  
  +\left(I_7^{(yxx)} p_c\cdot p_w-I_7^{(yxy)} p_c\cdot p_v\right)
  \frac{E_b}{m_ak_b} 
\\
w_D(k_x,k_y,z_x,z_y) & = & 
  I_1^{(1)} \left[ \frac{(m_y^2-m_c^2)(m_y^2-m_c^2
        +2( p_a- p_x)\cdot( p_c- p_y)}{m_w^2} - \right.   
\nonumber
\\  & &  
  \left. 
  -( p_c+ p_y)^2+2( p_a- p_x)\cdot( p_c+ p_y)) \right]
\end{eqnarray}

The isospin factors resulting from the summation over all intermediate 
states in the diagrams of Tables \ref{TabSumphipi} and \ref{TabSumphirho} 
are listed in Table~\ref{TabIF}. 

\begin{table}[htb] 
\begin{center}
\begin{tabular}{lc} 
\hline
   Mechanism   &  Isospin factor 
\\ 
\hline
  $p\bar{p}(^3S_1) \to \pi\rho\pi \to \phi\pi^0$      &   1   
\\[2mm]
  $p\bar{p}(^3S_1) \to \rho\pi\pi \to \phi\pi^0$      &   $\frac{1}{3}$  
\\[2mm]
  $p\bar{p}(^1S_0) \to \pi\pi\pi \to \phi\rho^0$ (A)  &   $\frac{5}{\sqrt{3}}$ 
\\[2mm]
  $p\bar{p}(^1S_0) \to \pi\pi\pi \to \phi\rho^0$ (B)  &   $\sqrt{3}$  
\\[2mm]
  $p\bar{p}(^1S_0) \to \pi\pi\omega \to \phi\rho^0$   &   $\sqrt{\frac{2}{3}}$ 
\\[2mm]
  $p\bar{p}(^1S_0) \to \pi\omega\pi \to \phi\rho^0$   &   $-\sqrt{\frac{2}{3}}$ 
\\[2mm] 
\hline
\end{tabular} 
\end{center}
\caption{\label{TabIF} 
The isospin factors corresponding to the three-meson door--way 
mechanisms $p\bar{p}\to xyz \to \phi X$.}
\end{table} 

\renewcommand{\theequation}{C\arabic{equation}}
\setcounter{equation}{0}
\section{Re-evaluation of $p\bar{p}\to\rho\rho\to\phi\pi$}
\label{pprhorho}

\begin{figure}[h] 
\begin{center}
(a) \hspace*{65mm} (b) \\[-0.5\baselineskip]
\mbox{
      \mbox{\epsfysize=65mm\epsffile{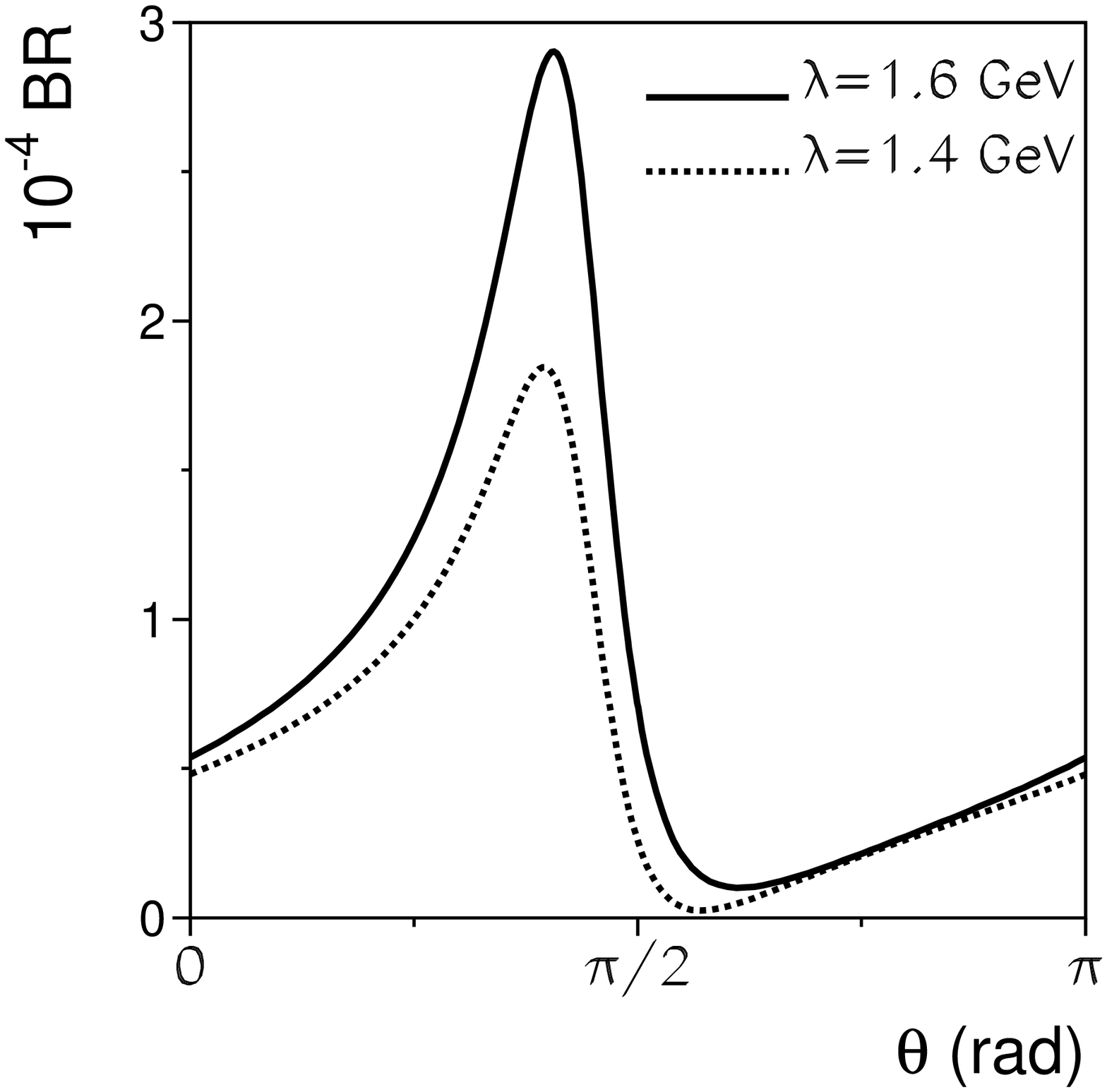}} \hspace*{1cm} 
      \mbox{\epsfysize=65mm\epsffile{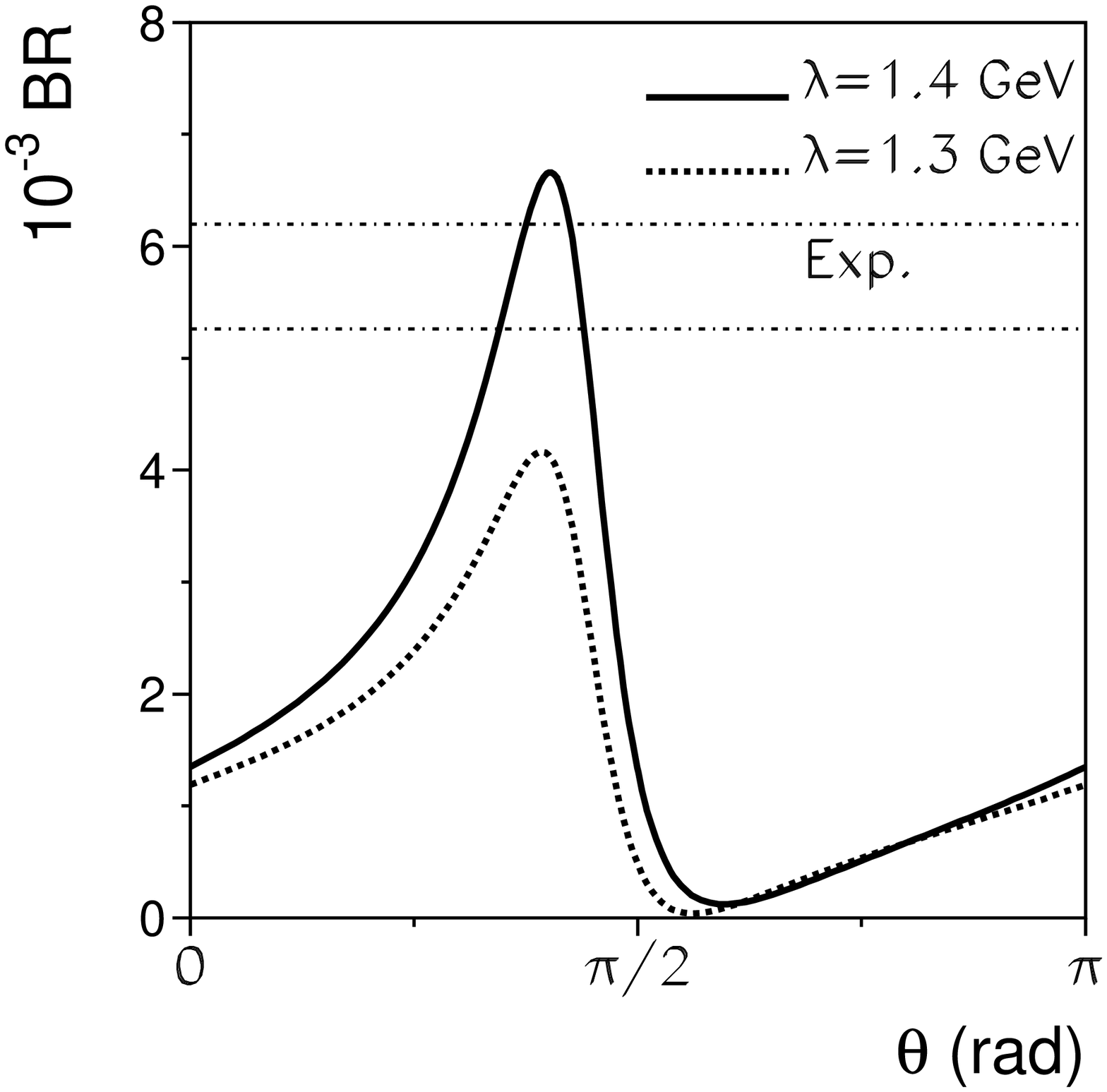}}             
} 
\end{center}\vspace{-1\baselineskip} 
\caption{\small The branching ratios corresponding to the door--way 
$\rho\rho$ mechanism for (a) $\phi\pi^0$ and (b) $\omega\pi^0$
production from $p\bar{p}(^3S_1)$ annihilation. The parameter $\vartheta$ defines
the relative strength of the two couplings in the $p\bar{p}\to\rho\rho$ vertex,
see Eqs. \protect \ref{T1} and \protect \ref{T2}.}
\label{BrfBro}
\end{figure}

For the $\rho\rho$ diagram the effective Langrangian is not unique
since the annihilation vertex $p\bar{p}\to\rho\rho$ allows for two invariant 
couplings with minimal number of derivatives. 
As in \cite{2MDM1}, these are denoted by
\begin{eqnarray}
T^{(1)}_{\pp\to\rho\rho} & = & g_1 \left( \ve_{\pp} \!\cdot\! \ve_x\,k_{\pp}
\!\cdot\! \ve_y-\ve_{\pp} \!\cdot\! \ve_y\,k_{\pp}\!\cdot\!\ve_x\right)
\label{T1} \\
T^{(2)}_{\pp\to\rho\rho} & = & g_2\,\ve_{\pp}\!\cdot\! k_x\,\ve_x\!\cdot\!\ve_y
\label{T2}
\end{eqnarray}
where $k_x$ and $k_y$ are the four-momenta of particles $x$ and $y$ and
$\ve_{\pp}$, $\ve_x$ and $\ve_y$ are the polarization vectors of the
corresponding particles. In \cite{2MDM1} the two cases were calculated
separately. Here we consider the coherent sum of both amplitudes 
\begin{eqnarray}
 T_{\pp\to\rho\rho} & = &  
 T^{(1)}_{\pp\to\rho\rho} \cos{\theta} +  T^{(2)}_{\pp\to\rho\rho} \sin{\theta}  
\end{eqnarray}
where the total strength is normalized to
$B\!R(\pp\to\rho\rho)=2.4\;\%$ from the theoretical estimate in \cite{2MDM0}.
Figure~\ref{BrfBro} shows the result of the calculations of the branching ratio 
for the $\phi\pi^0$ final state (including finite width effects) for the   
full space of parameters $g_1$, $g_2$.  To check the self consistency of 
the door--way calculations the contribution of the $\rho\rho$ intermediate 
state to the $\omega\pi^0$ production has been evaluated as well.  Since the 
$\omega\pi^0$ channel is not suppressed on the tree-level, it is gratifying  
that the one--loop corrections turn out to be small in comparison with the 
experimental data for this channel.  At the same time, the contribution to the 
OZI suppressed channel $\phi\pi^0$ is very significant for a broad range of 
the relative strength of  $g_1$ and $g_2$.  

\end{document}